# Durabilité sociale par l'engagement dans le contexte d'une formation avec des instruments tels que le Podcast natif et le réseau social Facebook

**Danielle BEBEY**

   Membre associé Laboratoire DICEN-IDF France
   LinkedIn :@daniellebebey

## Résumé

*La dimension sociale de la durabilité semble avoir été une notion rarement abordée par la littérature (Dubois et al., 2001) jusqu'au début des années 2000. A travers le colloque EUTIC 2023 il est donné la possibilité de se saisir de ce sujet de nouveau d'actualité. A cet effet, nous présentons un processus d'engagement qui s'inscrit dans une dynamique de développement durable en s'appuyant sur des instruments numériques inspirés de la vie quotidienne, pour des applications dans le contexte de la formation, dans un objectif d'apprentissage tout au long de la vie. Issu des sciences de l'information et de la communication, nos travaux se nourrissent d'un ancrage pluridisciplinaire qui selon nous, peut faire écho dans diverses disciplines, mais qu'il est intéressant de challenger, d'où l'objet de cette contribution.*

Mots-clés : engagement, formation pour adultes, podcast, réseau social numérique, Facebook.

## Abstract

*The social dimension of sustainability seems to have been a notion rarely addressed in the literature (Dubois et al., 2001) until the early 2000s. The EUTIC 2023 symposium provides an opportunity to take up this topical issue. To this end, we are presenting an engagement process that is part of a sustainable development dynamic, based on digital tools inspired by everyday life, for applications in the context of training, with a view to lifelong learning. Our work, which stems from the information and communication sciences, is rooted in a multi-disciplinary approach that we believe can be echoed in a variety of disciplines, but which it is interesting to challenge, hence the purpose of this contribution.*

Keywords : engagement, adult learning, podcast, digital social network, Facebook.

## Introduction

La problématique d'engagement en formation présentielle reste un enjeu majeur selon les analyses récentes (Hiver et al., 2021; Sitzmann & Weinhardt, 2018). Lorsque le numérique intègre les programmes, il donne l'impression de régler ce problème d'engagement à court terme. En revanche à long terme, la réalité est toute autre. Il persiste des interrogations sur les méthodes d'engagement en formation de manière générale, en présentiel comme à distance car les équipes pédagogiques ne sont pas toutes logées à la même enseigne en termes d'équipement et d'outils pédagogiques. La récente épidémie a fait resurgir ces questionnements non plus uniquement dans l'optique de tenter de trouver des solutions ponctuelles, mais d'être en capacité de les anticiper parce que les actions menées devraient s'inscrire sur la durée. C'est dans cet état d'esprit que nous avons réalisé nos travaux de thèse de doctorat. Il s'agissait de vérifier une hypothèse selon laquelle des dispositifs complémentaires à valeur expérientielle peuvent permettre d'accroître la participation des sujets pour consolider l'engagement sur la durée de la formation. Notre objectif était de développer des mécanismes d'apprentissage que les apprenants pourraient mobiliser de façon autonome, en situation réelle après la formation.

Pour ce faire, nous nous sommes lancés dans une recherche-action au cours de laquelle nous avons mis en place puis animé les dits dispositifs sous la forme d'expérience. L'objectif était d'observer les effets de ces dispositifs en suivant une approche mixte de collecte de données pour vérifier que des formes d'engagement pourraient s'en dégager. Les résultats obtenus permettent de réaliser que nous sommes dans un processus de développement durable qu'il peut être intéressant de transférer dans d'autres contextes pour jauger des limites et des apports à plus grande échelle, sur la durée et suivant l'évolution des usages des dispositifs.

Dans ce document, nous présenterons donc les théories que nous avons mobilisées pour ce travail tant sur les questions de durabilité sociale, d'engagement, et d'usabilité de dispositifs numériques pour un développement durable social et humain. Dans ce développement nous nous concentrons uniquement sur deux de ces dispositifs qui interpellent les humanités numériques. Il s'agit du *podcast* et du réseau social numérique Facebook.

## Revue de littérature

La durabilité sociale renvoie vers deux formes particulières de capital à savoir le capital humain et le capital social. « Le capital humain, sous les formes de capital éducatif et de capital santé, retrace les effets sur la personne d'une éducation suivie, de l'expérience professionnelle, et d'un suivi médical et nutritionnel adéquat » (Dubois et al., 2001, p. 9). « Le capital social « potentiel » a trait aux relations et interactions qui existent entre individus sous forme familiale, de bon voisinage, d'appartenance à des réseaux ou associations, de partage de normes et valeurs communes » (Dubois et al., 2001, p. 9).

Une analyse de la durabilité nécessite d'identifier les différentes dimensions du phénomène considéré et de mettre en place des indicateurs capables d'en mesurer les interactions (réactions, présence sociale, etc.). Pour ce faire, la littérature (Dubois et al., 2001) recommande de prendre en compte diverses dimensions et d'analyser également les interactions qui en découlent. Ainsi, pour qu'il soit considéré comme un processus de développement socialement durable, il faudrait que les acquis sociaux ne soient remis en question ni par les générations actuelles ni par les générations futures. Dans le cas où l'humain est interpellé, il s'agit particulièrement de s'intéresser au capital humain (éducation, santé) et au capital social (relations sociales) pour analyser le processus de durabilité.

Une notion semble faire écho à cette idée d'analyse de la durabilité et de mise en place d'indicateur. Il s'agit de l'engagement, une notion complexe qui nécessite de présenter le prisme de sa perception à chaque prise de parole. A cet effet, nous aimons bien cette définition de l'engagement donnée par Kiesler et Sakumura (p. 349) et relevée par (Bernard, 2012, p.20). Pour eux, « l'engagement est le lien qui unit l'individu à ses actes comportementaux ». Si la définition est claire quant à l'association de l'engagement à l'individu/l'humain, il faudrait lire entre les lignes pour comprendre que les actes comportementaux d'un individu s'apprécient dans un contexte, un environnement et potentiellement dans des interactions sociales. Ainsi, dans un contexte de gestion et management, l'engagement peut tenir compte des évolutions permanentes, de la complémentarité des individus et de leurs différentes appartenances (Damart, 2006). Dans un contexte d'éducation, l'engagement peut être perçu comme « l'implication active et volontaire, manifestée par une multitude d'actes de décision interactifs et complémentaires qui structurent et sont structurés par la conduite vis-à-vis de la formation, en vue de la réalisation d'un projet personnel » (Kaddouri, 2011, p.75). Cette notion s'associe donc facilement au capital humain et social de la durabilité sociale.

Ainsi, en choisissant d'observer l'engagement dans le cadre de la durabilité sociale, il convient d'identifier les indicateurs associés. En parcourant la littérature (Bonfils & Ghoul Samson, 2018), nous constatons que l'observation de l'engagement du point de vue de l'expérience sur un long terme, s'est principalement effectuée dans des espaces numériques dans de nombreuses disciplines. Les humanités numériques représentent donc un savoir construit, manipulable à la fois par l'homme et la machine suivant une méthode et un résultat aboutissant à une transformation d'usage (Berra, 2012). Il se dessine une piste d'observation de l'engagement par l'expérience qui prend en compte les notions de durée. Sachant que les limites des enquêtes classiques relevées (Dubois et al., 2001) portent sur leur manque de prise en compte du temps et de la durée, nous y voyons un intérêt. En ce qui concerne les indicateurs d'appréciation de la qualité des expériences, O'Brien et Toms (2008) ont présenté quelques-uns à savoir l'attention suscitée, les effets positifs, la diversité des propositions, le contrôle perçu, la rétroaction, l'interactivité, l'attrait esthétique et/ou émotionnel. Pour Holbrook & Hirschman (1982), la valeur expérientielle peut également être mesurée à travers la satisfaction d'un point de vue symbolique, hédonique ou esthétique. Ces indicateurs semblent contribuer à transformer l'expérience vécue par les participants et ne la limitent plus à « un processus linéaire mais comme un processus en boucle qui se transforme en fonction des espaces parcourus, mais aussi de la culture médiatique, des perceptions et des actions sous différentes formes de corporéités du sujet » (Bonfils, 2014, p. 28). Au travers de ces indicateurs, nous percevons une opportunité d'analyser la durabilité sociale par l'engagement.

Dans la littérature de Dickey (2005) nous identifions des points de convergence entre les théories de l'engagement et l'approche constructiviste que nous pouvons mobiliser en sciences de l'information et de la communication. Nous y voyons une piste d'observation de l'engagement dans notre cas d'étude, au sens de Maillet et Lemoine (2007). Il s'agit d'observer l'engagement dans des contextes spécifiques où les formes de participation sont ouvertes à la diversité et cette diversité est ancrée dans les générations. Dans la même veine, Jouët (1993) a suggéré d'adopter des pratiques de communication diverses, originales et complémentaires inspirées de la vie quotidienne. Cardon (2006) conforte cette idée en recommandant de faire participer l'agent bénéficiaire directement à la conception d'innovations à partir des technologies et services mis à sa disposition pour un engagement profond. Notre principe d'observation de l'engagement est aligné avec ces recommandations dans une logique de durabilité sociale. Les résultats de l'observation doivent donc faire ressortir les effets des interactions dans le temps et sur la période pour les générations actuelles et futures.

Un cadre d'observation semble se prêter à cette logique de durabilité sociale. Il s'agit de l' « école ». Elle est perçue comme un laboratoire d'expérimentation (Lange, 2015) qui ne se réduit pas à adapter l'apprenant à l'actuelle société des adultes, mais devance et anticipe toujours pour faire vivre au présent les enjeux d'une action collective et collégiale essentielle. L'Ecole est donc structurée en prenant compte des modes d'enseignement plus fondamentaux qui ont des effets sur la durée, et le numérique constitue un puissant vecteur d'une démocratisation de la connaissance. L'engagement lors de l'apprentissage s'inscrit de cette manière dans une dynamique de développement durable. Ainsi, de nombreux voyants semblent être au vert pour observer la durabilité sociale par l'engagement en formation avec des indicateurs de mesure des interactions et des instruments numériques.

## Approche et méthodologie de recherche

La littérature de Dubois et al. (2001), nous interpelle sur les questions de mesure des interactions pour observer la durabilité sociale par l'engagement. Elle propose de mobiliser des instruments « particuliers » car les enquêtes statistiques classiques ne semblent répondre que partiellement à l'élaboration des indicateurs. Dubois et al. (2001) ne se sont pas contentés de présenter les limites des solutions existantes. Ils mentionnent aussi des enquêtes qui pourraient être plus efficaces. Ils parlent notamment d'enquêtes à passage répétés, de systèmes d'enquête, d'enquêtes permanentes, et plus récent, les enquêtes à phase avec une répétition d'enquêtes plus ou moins articulées. L'idée de cette répétition serait de produire régulièrement de l'information, sous quelque forme que ce soit sur un phénomène donné. L'enrichissement perpétuel a pour but d'ajuster le recueil et la production d'information en fonction des besoins. Cet enrichissement permet également de combiner différents instruments de collecte et par la même occasion de fournir des indicateurs. Sur la base de ces principes, les observations peuvent s'effectuer soit à travers le suivi d'une population sur une période spécifique, soit à travers le suivi d'une thématique particulière via diverses sources de données.

Par ailleurs, Tricot (2015) souligne que « c'est pas l'analyse de la tâche uniquement qu'il faut prendre en considération c'est l'analyse de la tâche et le mode d'engagement dans la tâche ». Cette recommandation nous pousse également à nous interroger sur une méthode d'observation de l'engagement qui suivrait une approche plus globale. La démarche par recherche-action participative (Gonzalez-Laporte, 2014) suivant la théorie ancrée semble correspondre à ce modèle d'observation. Elle permet d'établir des aller-retours entre la littérature, le terrain, nos réflexions et tirer le maximum de notre étude exploratoire. Ainsi, le potentiel du terrain a été appréhendé à travers une recherche mixte (Creswell ; Creswell, 2017). Elle nous permettait d'allier l'approche quantitative descriptive et l'approche qualitative ethnographique multi située (Marcus, 1995) qui autorise la collecte sur divers échantillons en lieu et contexte différents pour enrichir les précédents résultats. De plus, elle permettait de suivre une population sur une période spécifique, mais également de suivre notre thématique sur l'engagement via diverses sources de données.

Nous suivons les recommandations de Jouët (1993) et Cardon (2006) en identifiant des dispositifs potentiels inspirés de la vie quotidienne, qui laisseraient place à une forme de participation de l'agent bénéficiaire. Sachant que l'aspect utilitaire influence la pérennité des usages (Andonova, 2009), l'objectif est de confier des tâches faciles à réaliser et à faible valeur ajoutée comme proposé par d'autres chercheurs (Mencarelli & Rivière, 2014), pour susciter un engagement. Pour ce faire, nous avons identifié des tendances d'usage qui pourraient favoriser l'engagement des bénéficiaires. Ainsi, nous nous sommes aperçus que le *podcast*[1] a largement été utilisé dans l'enseignement (Dale, 2007). A travers ce dispositif, les apprenants peuvent participer au processus de production de la ressource numérique et mobiliser des connaissances antérieures pour un projet commun. De plus, le dispositif favorise la mémorisation par la répétition en formation d'après l'analyse de McGarr (2009).

Par ailleurs, le réseau social Facebook a été identifié comme étant un outil qui favorise l'amélioration de connaissance durant l'apprentissage (Aubert & Froissart, 2014). Il semble être très efficace sur les questions d'interactions sociales (Nguyen & Lethiais, 2016). L'usage de Facebook est corrélé à la satisfaction personnelle, à une meilleure confiance en les autres, et un engagement plus significatif dans des actions sociales et collectives. Les réseaux sociaux plus particulièrement Facebook et twitter, contribuent à l'expérimentation des humanités numériques. Ils permettent de tracer et d'observer l'identité et la dynamique de médiation d'une communauté de façon asynchrone. Ces dispositifs offrent des possibilités d'observation de l'engagement en formation à travers des indicateurs de satisfaction entre autres. Les deux expériences ont été choisies dans le but d'offrir une dimension authentique (Herrington et al., 2014) à la formation présentielle afin de la prolonger en dehors de l'espace présentiel de cours. La spécificité de l'apprentissage authentique est qu'elle traite des problèmes complexes du monde réel et leurs solutions à travers une approche pédagogique qui projette vers une utilisation future. L'approche pédagogique que nous adoptons dans ces travaux suit également une logique de durabilité sociale en mobilisant à travers ces dispositifs le capital humain avec la formation et le capital social avec des expériences interactives qui nécessitent des implications des bénéficiaires.

Pour l'analyse des données issues du travail effectué avec ces dispositifs, nous avons opté pour une triangulation entre les échanges informels et flux de données sur les groupes Facebook, l'entretien de groupe et le questionnaire soumis après chaque expérience. L'étude du terrain par le biais de plusieurs méthodes adoptées simultanément, concourt à la bonne gestion du temps de collecte. Les résultats que nous présentons, font état d'une approche globale qui ne se limite pas simplement aux deux expériences. Nous tacherons cependant de présenter quelques retours spécifiques à ces dispositifs numériques pour faciliter la projection du lecteur. Les détails de cette démarche peuvent être analysés en profondeur dans nos travaux de thèse ou les autres publications associées. Nous avons établi une triangulation simple grâce à la collecte sur un échantillon spécifique et une triangulation par différentes techniques pour enrichir notre étude par une complémentarité de sources ; ce qui permet par la même occasion d'approfondir la compréhension de l'objet d'étude. En plus de notre revue de littérature, notre approche terrain semble également correspondre au principe de durabilité sociale de par la diversité des méthodes de collectes et l'intégration d'enquêtes répétées.

---

[1] Podcasting = Web syndication (RSS, Atom) + Audio content (talk-shows, music, news, and certainly learning resources…) + Mobile devices (mp3 players, PDAs, cell phones…). (Cebeci & Tekdal, 2006)

L'expérience sur Facebook s'est tenue durant l'année académique 2017-2018 sur 4 mois. Celle avec le *podcast* s'est déroulée lors d'une séance de 3 h, l'ensemble avec 21 participants âgés de 18 à 45 ans et vivants en Île-de-France. Une des unités d'enseignement s'est tenue d'octobre 2017 à février 2018 pendant 30h pour le compte du premier semestre de la formation professionnelle intitulée pratiques de communication et de négociation. Ce cours dispensé en cours du soir, était à destination d'adultes travailleurs ou en recherche d'emploi pour le poste de chargé d'accompagnement social et professionnel. Les 7 participants à cette unité d'enseignement avaient un âge compris entre 25 et 45 ans. En parallèle, nous dispensions un cours d'expression et communication à 14 jeunes en DUT 2$^{ème}$ année génie électrique par alternance. Les DUT avaient un âge compris entre 18 et 25 ans. Leur formation s'est déroulée tout au long de l'année 2017-2018 sur 2 semestres de 30h au total. A la suite de chaque unité d'enseignement, suivant le principe de la théorie ancrée, nous tirons des enseignements et travaillons sur des aspects différents avec le groupe suivant. Cette approche permet de suivre la dynamique de durabilité sociale sur la période de recherche car chaque groupe constitue une sorte de génération et les informations acquises avec un groupe permettent d'enrichir le nouveau (la nouvelle génération) et également de collecter leurs retours d'expérience.

Nous avons saisi l'occasion de la proposition d'expérience par le *podcast* pour détourner l'entretien de groupe pour nos travaux. Ainsi, les apprenants devaient constituer une forme d'émission radio suivant un cadrage spécifique en s'appuyant sur leur formateur, leurs connaissances antérieures et leur créativité. Nous avons suivi cette même logique sur le groupe classe Facebook. Cependant, la thématique de ce *podcast* à diffusion restreinte aux groupes, bien que proposée dans le but de pratiquer l'argumentation, permettait également de compléter notre collecte de données. Le caractère natif du *podcast* s'explique par le fait qu'il était diffusé en ligne au groupe classe sur Facebook et le groupe était « fermé » aux participants de la formation uniquement. Sur Facebook, nous avons animé 2 petites communautés dans le cadre des 2 formations pour montrer une possibilité d'exploitation des outils à disposition tels que les réseaux sociaux pour l'apprentissage. L'objectif était d'accompagner les apprenants de manière à observer une évolution individuelle et collective au moins sur la période de formation.

## Résultats et interprétations

Nous avons choisi d'observer l'engagement en formation à travers des expériences spécifiques. Plusieurs indicateurs d'observation ont été proposés par la littérature. Nous avons pris le parti, avec le temps imparti pour les travaux, de traiter l'engagement principalement autour d'indicateurs de satisfaction sur le plan utilitaire et prospectif.

D'un point de vue qualitatif, les retours des apprenants sur les expériences proposées permettent de matérialiser la durabilité au-delà de l'expérience spécifique, car les apprenants prennent conscience des connaissances acquises, de leurs progrès et de potentielles exploitations en dehors du cadre de formation. Les verbatims suivants nous le présentent: « faire un débat et utiliser ce qu'on a appris permet de bien assimiler et les projets comme la radio nous permettent d'improviser, nous impliquer plus, et que ça soit plus intéressant » ; il s'agissait pour eux d'« un exercice nouveau et ludique qui mettait à contribution les qualités de réflexion et d'observation de tous les élèves » ; ils ont considéré que « les mises en pratique permettent d'aller à l'essentiel du cours et facilitent la mémorisation » ; pour certains « l'expression de mes idées est plus structurée je trouve, j'ai pu le remarquer lors des débats mais aussi en dehors de la classe » ; « ce semestre j'ai plutôt échangé avec mes camarades de classe… J'ai beaucoup apprécié l'exercice web radio, nous avons pu discuter et échanger même avec des opinions différentes » ; « le débat que l'on a fait m'a permis de m'exprimer sans stress » ; « j'ai progressé pour la structuration de mes écrits et oraux. J'essaie d'aller à l'essentiel et de donner les informations importantes quand il faut être bref » ; « je pense avoir fait des progrès significatifs par rapport au début de l'année ». Avant la classe virtuelle, l'utilisation du réseau social Facebook pour cet échantillon se limitait à un usage privé et réduit. En définitive, l'expérience s'est avérée être une « autre manière d'apprendre », de rompre avec le format « distanciel classique », d'apprendre « plus vite », de « voir sa progression », de « rapprocher les auditeurs », une « découverte intéressante ».

D'un point de vue quantitatif, sur la perception de l'expérience, nous mettons en avant les éléments qui ont suscités un intérêt pour les bénéficiaires. Les apprenants ont évoqué « la possibilité de s'exercer sous forme de projet » dans le cas du *podcast*, et ont également relevé l'ambiance et le travail en groupe. Quelques-uns (5) ont notifié que la séquence de *débriefing* et la réécoute du *podcast* pour leur amélioration a été déterminante. L'observation des interactions Facebook avec l'outil Grytics permet de noter une évolution progressive de l'activité à travers les commentaires, les réactions, les *posts*. Nous avons observé un score d'engagement qui évolue positivement au fil du temps avec un taux de 75,31 %. L'interaction s'est matérialisée par l'échange d'informations, la manifestation d'affects (commentaires, réactions) entre les intervenants. 90 % (19/21) des répondants ont affirmé que les dispositifs proposés ont fortement contribué à la mémorisation du cours. 30 % (6/21) ont approuvé la situation avec une note de 10/10. 25 % (5/21) ont attribué une note de 8/10. Ce qui veut dire que pour cet échantillon, les expériences ont constitué pour une bonne majorité un indicateur clé de succès dans la mémorisation des notions abordées durant les séances pratiques de formation. Il s'est développé de nouveaux comportements qui caractérisent l'engagement sur la durée de la formation.

## Conclusion et discussion

Notre développement présente les résultats de la mesure de la durabilité sociale par l'engagement en formation. Il s'agit de résultats d'un processus coconstruit qui s'étend sur la durée via le numérique dans le contexte de la formation pour adultes. L'objectif de cette étude exploratoire était de donner les billes aux parties prenantes pour être progressivement autonome en situation réelle. Nous sommes partis du postulat selon lequel les expériences coconstruites avec les apprenants dans le but de les engager dans la consommation de leur formation sur le long terme, favorisaient leurs participations ponctuelles et permettaient progressivement d'arriver à ce résultat positif qu'est l'engagement. Pour ce faire, nous avons détourné les usages de 2 instruments déjà populaires dans le quotidien. Il s'agit d'une part d'une expérience de formation via un groupe classe sur Facebook et de l'autre, d'un mini projet autour de la mise en place d'un *podcast* natif.

L'analyse des informations dans le cadre d'une recherche-action permet de présenter les résultats obtenus. L'hypothèse sur l'engagement est validée sur la période de formation avec cet échantillon et certains indicateurs laissent penser qu'il pourrait y avoir un avis favorable quant à la poursuite d'utilisation de ces dispositifs si l'occasion leur était donnée. L'intérêt d'un apprentissage tout au long de la vie est la possibilité d'adapter, approfondir, rappeler. Nous avons conscience que sans pratique, après quelques mois, ces notions peuvent être perdues et il peut être nécessaire de faire une piqure de rappel. Les perspectives de ces travaux seraient d'observer ce phénomène à travers une étude longitudinale pour renforcer la proposition. Néanmoins les résultats obtenus, démontrent qu'il s'est développé de nouveaux comportements qui caractérisent l'engagement dans un apprentissage authentique, qui fait écho sur les principes de durabilité sociale.

L'incomplétude de notre représentation du terrain du fait que nous soyons impliqués dans la collecte de données, rend déjà notre analyse partiellement limitée. La posture de jeune chercheur que nous avions à cette période, pourrait aussi justifier certaines limites dans notre analyse, tout comme l'échantillon restreint. Cette approche se concentre sur la satisfaction expérientielle de l'utilisateur et la plupart des données sont déclaratives. Ce qui peut constituer une autre limite de ce travail. Cependant, nous suivons une démarche scientifique qui permet d'identifier notre ancrage théorique, notre positionnement, notre méthodologie et quelques perspectives associées. Cette étude permet de mesurer et valider l'hypothèse concernant l'engagement en formation par recherche-action. Nous espérons que les éléments partagés dans ce document permettent de visualiser les indicateurs de durabilité sociale par l'engagement à travers des aspects rationnels (analyse, logique) et des aspects irrationnels (émotionnels, sociaux) observables. Ces éléments permettent de démontrer en quoi la proposition de tels instruments sur des scénarios spécifiques, dans un objectif d'engagement pourrait contribuer à la durabilité sociale en formation.

L'enseignement que nous tirons de ces travaux sur ce sujet, est que pour qu'on puisse parler d'une durabilité sociale, il faudrait impliquer les bénéficiaires, leur proposer des formes de participation et de collecte diverses, flexibles et complémentaires qui s'inscrivent dans l'ère du temps pour une validation des différentes générations. Ces propositions, pour être efficaces, doivent également tenir compte des attentes, besoins, appétences et sensibilité des parties prenantes tout en offrant la possibilité de progresser (donner du sens individuellement et collectivement). La durabilité sociale dans notre cas s'observe à plusieurs niveaux, notamment à travers la méthode de collecte, les dispositifs mobilisés et les indicateurs de mesure de l'engagement. A travers la méthode exploitée, nous pouvons dire que l'échantillon observé a validé les exploitations actuelles et futures de notre proposition sur la durée de la formation. Les dispositifs mobilisés ont probablement favorisé cette situation du fait qu'ils soient déjà installés dans le quotidien et qu'ils aient fait leur preuve en formation. En serait-il de même quelques années plus tard ? En serait-il de même sans les outils numériques ? ces outils spécifiques ? Les réponses à ces interrogations pourraient faire l'objet d'autres travaux dans le but d'enrichir la réflexion.

## Bibliographie